# Life history evolution and the origin of multicellularity: differentiation of types, energy constraints, non-linear curvatures of trade-off functions

Denis Tverskoy[1]

## Abstract

A fundamental issue discussed in evolutionary biology is the transition from unicellular to multicellular organisms. Here we develop non-robust models provided in [1] and attempt to get robust models investigated how differentiation of types and energy constraints influence on the optimal behavior of colonies with different size (so, different initial costs of reproduction).

Constructed models show that each large - sized colony with high initial costs of reproduction tends to full specialization, no matter are all cells in this colony identical or are there cells with different types in this colony. Moreover, in this model (as distinct from [1]) not exactly the half of cells from the colony should specialize in, for example, soma. The level of type's diversity determines the number of cells specialized in soma. In small – sized colonies with low initial costs of reproduction, when type's diversity is week, an unspecialized state may bring colony some benefits. However, these benefits may be only local and in optimum in the colony some cells would be specialized, others – unspecialized. The amount of specialized cells in small – sized colony depends on the level of type's diversity in this colony.

Adding energy constraint, we may receive robust models even in convex case. In optimum, the colony with different types of cells and energy restriction may be indifferent between some optimal patterns of states. Arbitrary chosen cell may be soma or germ in some states or may be unspecialized in other. Moreover, in each optimal state levels of fecundity and viability of each cell lie in limited ranges. This result reflects the fact that some cell in the colony may lose the potential ability to achieve, for example, high level of fecundity, but does not lose the possibility to perform a reproductive function at all. It means that provided model can describe organisms, which represents the intermediate between unspecialized colonies and full-specialized multicellular organisms.

Also, constructed full optimization model with different type of cells and energy restrictions reveals an important property: a colony of cells in some cases may reallocate (without any loses in fitness of the colony) fecundity and viability between cells in response to corresponding displacement of different external irritants.

## 1. Introduction

---

[1] disa1591@mail.ru, International Laboratory for Decision Analysis and Choice and Department of Mathematics for Economics of the National Research University Higher School of Economics (NRU HSE), and Institute of Control Sciences of Russian Academy of Sciences

In [1] fitness of a colony is considered in terms of its two basic components: viability and fecundity. Authors show that the group fitness trade-offs are initially determined by the cell level trade-offs. So, in [1] is suggested that the curvature of trade-off functions is an important factor influenced on the process of unicellular – multicellular transition. Moreover, [1] predicts that small-sized colonies with low initial costs of reproduction have concave trade-off functions of each cell; large – sized colonies require high initial costs of reproduction and hence, convex trade-off functions. Beautiful models provided in [1] look like non-robust (see [2]). In [2] authors try to develop model from [1] in order to receive a robust one. Also they investigate the 'pure' impact of environmental factors and diversity of cells' types on colony's incentives to specialization. In this work we attempt to develop models from [1] in order to get robust models, which allow to investigate the impact of factors from [2] on the colony's well – being, depending on the size of the colony under consideration.

**Structure of the text**

We begin with full optimization model with different types of cells and convex trade-off functions and describe it in Section 2. Then in Section 3 we propose the model with different cell types and concave trade-off functions. We extend these models taking into account energy constraints in Section 4. Section 5 concludes.

## 2. Full optimization model with different types of cells: the case of convex trade-off functions

### 2.1. Formulation of the problem

Consider the full optimization problem with different types of cells provided in [2]. In this model each cell may have its own type, reflected in the intrinsic trade-off function of this cell. Thus, the diversity in types of cells corresponds to the diversity in two parameters of linear trade-off functions. Linearity in this model represents a very convenient assumption, because it allows to understand easily the essence of the solution and to reveal main properties of this solution. But the case of linear functions cannot be common enough. It makes sense to propose that the shape of trade-off functions may be complicated in some cases. In this paragraph we suggest a model with convex trade-off functions. Convex trade-off function describes the principle of the decreasing marginal rate of substitution: in case of a little fecundity, in order to increase the fecundity per unit, cell should abandon more than the unit of viability; in situation when the cell achieves a big level of fecundity, each additional unit of it requires less than one unit of declining viability. Moreover, this type of trade-off functions corresponds to large size colonies with high level of initial costs of reproduction. So we investigate the impact of diversity of types on the large size colony's well-being.

In our model, the diversity of types of cells corresponds to the diversity of convex functions; each of them represents the trade-off relation within some cell from the colony.

Consider the group from N cells.

$i = 1 \ldots N-$ indexes of cells in the colony.

$b_i$ - the level of fecundity of cell i.

$v_i$ – represents viability of cell i.

The group's level of fecundity is an additive function of variables $b_i$. The group's level of viability is an additive function of variables $v_i$:

$$B = \sum_{i=1}^{N} b_i, \quad V = \sum_{i=1}^{N} v_i. \tag{1}$$

The group fitness function (2) reflects the synergetic effects of jointly existing cells into the colony:

$$W = V * B. \tag{2}$$

Consider now trade-off functions of cells:

$$\forall i = \overline{1,N}: v_i = \varphi_i(b_i). \tag{3}$$

Also, we provide a set of conditions, which restrict the shape of intrinsic trade-off function of each cell:

1. for any $i = \overline{1,N}$ let $b_i^{max} \in R, 0 < b_i^{max} < \infty$;

2. $\forall i = \overline{1,N}: \varphi_i: [0, b_i^{max}] \to R, \varphi_i \in \mathbb{C}^2_{[0, b_i^{max}]}$;

3. $\forall i = \overline{1,N}: \varphi_i(0) = v_i^{max}, 0 < v_i^{max} < \infty$;

4. $\forall i = \overline{1,N}: \varphi_i(b_i^{max}) = 0$;

5. $\forall i = \overline{1,N}: \frac{d\varphi_i(b_i)}{db_i} < 0, \forall b_i \in [0, b_i^{max}]$;

6. $\forall i = \overline{1,N}: \frac{d^2\varphi_i(b_i)}{db_i^2} \geq 0, \forall b_i \in [0, b_i^{max}]; \exists j \in \{1..N\}: \frac{d^2\varphi_j(b_j)}{db_j^2} > 0, \forall b_j \in [0, b_j^{max}]$.

Assumption (2) embodies the idea about a quite simple nature of trade-off functions and allows us to apply common techniques in order to get a solution to the model. (1), (3) and (4) prohibit cells to achieve the infinite level of fecundity or viability and fix boundary conditions for each trade-off function. Next assumption (5) maintains the biological idea of a decreasing trade-off function: the bigger level of fecundity corresponds to the bigger reproductive effort of cell, which causes a less level of viability. (6) shows that viewed functions are convex. Also we should note that, in common, all trade-off functions may be different. In particular, they all may be identical. This fact means that provided model generalizes corresponding model in [1] – full optimization model with identical cells in the colony and convex trade-off functions.

So, we can construct provided model as a classic optimization problem with constraints:

$$\begin{cases} W = \sum_{i=1}^{N} b_i * \sum_{i=1}^{N} v_i \to max_{b,v} \\ \forall i = \overline{1,N}: v_i = \varphi_i(b_i), \\ \forall i = \overline{1,N}: b_i \geq 0, \\ \forall i = \overline{1,N}: v_i \geq 0. \end{cases} \leftrightarrow \begin{cases} W = \sum_{i=1}^{N} b_i * \sum_{i=1}^{N} \varphi_i(b_i) \to max_b \\ \forall i = \overline{1,N}: 0 \leq b_i \leq b_i^{max}. \end{cases} \quad (4)$$

Generally speaking, we should maximize the twice continuously differentiable function, determined in a compact set in $R^n$ space and analyze the solution to this problem in order to reveal the optimal behavior of each cell within the group.

## 2.2. Analysis of the model

Consider task (4). Choose arbitrary and fix parameters of this task according to assumptions (1)-(6). In this section we would consider only this chosen problem as the problem (4). Because we chose this task arbitrary, all inferences about the solution to this problem would hold to each task (4).

Further in this section we would construct auxiliary optimization problems based on parameters of chosen task (4). These tasks help us to solve optimization problem (4) and infer more about its solution.

**Definition 1.** Let $\tilde{\mathcal{P}}$ be the class of optimization problems. $p \in \tilde{\mathcal{P}}$ if and only if p is an optimization problem which form is represented below:

$$\begin{cases} W_p = \left( \sum_{i \in I_1^p} b_i^{max} + \sum_{i \in I_3^p} b_i \right) * \left( \sum_{i \in I_2^p} v_i^{max} + \sum_{i \in I_3^p} \varphi_i(b_i) \right) \to max_{b_i, i \in I_3^p} \\ \forall i \in I_3^p: 0 \leq b_i \leq b_i^{max}. \end{cases} \quad (5)$$

$$\text{where} \quad \begin{cases} I_1^p \subset \{1,..,N\}; I_2^p \subset \{1,..,N\}; I_3^p \subseteq \{1,..,N\}; \\ I_1^p \cup I_2^p \cup I_3^p = \{1,..,N\}; \\ I_1^p \cap I_2^p = I_1^p \cap I_3^p = I_2^p \cap I_3^p = \emptyset; \\ |I_3^p| \geq 1; \end{cases} \quad (6)$$

So, any task p from the class $\tilde{\mathcal{P}}$ has a form (5) and is determined exactly of a triple of sets $(I_1^p, I_2^p, I_3^p)$ satisfying to conditions (6). For example, problem (4) also belongs to class $\widetilde{\mathcal{P}}$ and is matched to the triple $(\emptyset, \emptyset, \{1,..,N\})$ (denote it as $\tilde{\mathcal{P}}\tilde{\mathcal{P}}$). Notice, that this definition is constructed by parity of reasoning with the definition 1, introduced in [2].

The system of notation, which is provided in [2], is very useful, so we remind it here, applying to the viewed model.

$F_p$ represents the domain of optimization task p, $p \in \tilde{\mathcal{P}}$

$$H_p = \{b \in R^N: 0 \leq b_i \leq b_i^{max}, \forall i \in I_3^p; b_i = b_i^{max}, \forall i \in I_1^p; b_i = 0, \forall i \in I_2^p\}, p \in \tilde{\mathcal{P}}$$

$$H = \{H_p, p \in \tilde{\mathcal{P}}\}$$

$$L = \{H_p, |I_3^p| = 1\}$$

$$F = \left\{b \in H_{\tilde{p}\tilde{p}}: \exists l \in L, \; b \text{ is a conditional stationary point of a task } \begin{Bmatrix} W(b) \to max \\ b \in l \end{Bmatrix}\right\}$$

**Theorem 1.**

Let $b^*$ be the solution to the problem (4). Then, the following statement is true:

$$\left(b^* \in Vert(H_{\tilde{p}\tilde{p}})\right) \vee (b^* \in F) \qquad (7)$$

The proof of this theorem you can find in Appendix A.

This theorem announces that in our model there would be no more than the one unspecialized cell. It means in biological terms that each large sized colony with high initial costs of reproduction tends to full specialization, no matter are all cells in this colony identical or are there cells with different types in this colony.

Moreover, in this model not exactly the half of cells from the colony should specialize in, for example, soma. The level of type's diversity determines the number of cells specialized in soma. Due to this fact we can claim that provided model overcomes some kind of non-robustness emerged in corresponding model in [1], but all in all is non-robust too:

- There can be situations when in solution to provided model only one cell remains unspecialized.

- When some cells in the colony are identical, we can get situations, when in the optimum one part of these cells should specialize in one function and another part – in another function. But due to identity of cells, it does not matter what cell exactly perform a particular function. So, the small deviation in unobserved parameters of the model can force soma- specialized cell become germ- specialized immediately.

3. **Full optimization model with different types of cells: the case of concave trade-off functions**

3.1. **Formulation of the problem**

Now we consider a full optimization problem which is differ from the previous one due to a concave shape of individual trade-off functions. In biological terms this model corresponds to a small- sized colony with low or without any initial costs of reproduction. As we know, small – sized colonies usually remain unspecialized. This fact is illustrated in model provided in [1], where all cells from a colony are identical. As we know from [2], the diversity in types is an important factor that tends colony to specialization. Our aim is to figure out how the diversity in types influences on the possibility of specialization in small- sized colonies. For this purposes we construct full optimization model with different types of cells and concave trade-off functions.

This model described by following optimization problem:

$$\begin{cases} W = \sum_{i=1}^{N} b_i * \sum_{i=1}^{N} v_i \to max_{b,v} \\ \forall i = \overline{1,N}: v_i = \omega_i(b_i), \\ \forall i = \overline{1,N}: b_i \geq 0, \\ \forall i = \overline{1,N}: v_i \geq 0. \end{cases} \leftrightarrow \begin{cases} W = \sum_{i=1}^{N} b_i * \sum_{i=1}^{N} \omega_i(b_i) \to max_b \\ \forall i = \overline{1,N}: 0 \leq b_i \leq b_i^{max}. \end{cases} \quad (8)$$

Where each intrinsic function satisfies to the set of assumptions (9):

1. for any $i = \overline{1,N}$ let $b_i^{max} \in R, 0 < b_i^{max} < \infty$;

2. $\forall i = \overline{1,N}: \omega_i: [0, b_i^{max}] \to R, \omega_i \in \mathbb{C}^2_{[0,b_i^{max}]}$;

3. $\forall i = \overline{1,N}: \omega_i(0) = v_i^{max}, 0 < v_i^{max} < \infty$;

4. $\forall i = \overline{1,N}: \omega_i(b_i^{max}) = 0$;

5. $\forall i = \overline{1,N}: \frac{d\omega_i(b_i)}{db_i} < 0, \forall b_i \in [0, b_i^{max}]$;

6. $\forall i = \overline{1,N}: \frac{d^2\omega_i(b_i)}{db_i^2} < 0, \forall b_i \in [0, b_i^{max}]$.

These assumptions are explained in the appropriate paragraph above; we only replace the suggestion of convexity by the suggestion of concavity. Also note that this model does not prohibit cells to have different trade-off functions. Therefore this model generalizes corresponding model provided in [1].

Our aim is to figure out circumstances under which the full specialization occurs in describing model.

### 3.2. Analysis of the model.

Consider task (8). Choose arbitrary and fix parameters of this task according to assumptions (9). In this section we would consider only this chosen problem as the problem (8). Because we chose this task arbitrary, all inferences about the solution to this problem would hold to each task (8) with any possible parameters satisfying to (9).

Further in this section we would construct auxiliary optimization problems based on parameters of chosen task (8). These tasks help us to solve chosen optimization problem (8) and infer more about its solution.

**Definition 2.** Let $\tilde{S}$ be the class of optimization problems. $s \in \tilde{S}$ if and only if s is an optimization problem which form is represented below:

$$\begin{cases} W_s = \left( \sum_{i \in I_1^s} b_i^{max} + \sum_{i \in I_3^s} b_i \right) * \left( \sum_{i \in I_2^s} v_i^{max} + \sum_{i \in I_3^s} \omega_i(b_i) \right) \to max_{b_i, i \in I_3^s} \\ \forall i \in I_3^s: 0 \leq b_i \leq b_i^{max}. \end{cases} \quad (10)$$

$$\text{where } \begin{cases} I_1^s \subset \{1,..,N\}; I_2^s \subset \{1,..,N\}; I_3^s \subseteq \{1,..,N\}; \\ I_1^s \cup I_2^s \cup I_3^s = \{1,..,N\}; \\ I_1^s \cap I_2^s = I_1^s \cap I_3^s = I_2^s \cap I_3^s = \emptyset; \\ |I_3^s| \geq 1; \end{cases} \quad (11)$$

So, any task s from the class $\tilde{S}$ has a form (10) and is determined exactly of a triple of sets $(I_1^s, I_2^s, I_3^s)$ satisfying to conditions (11). For example, problem (8) also belongs to class $\tilde{S}$ and is matched to the triple $(\emptyset, \emptyset, \{1,..,N\})$ (denote it as $\tilde{S}\tilde{S}$).

Introduce following notation:

$F_s$ represents the domain of optimization task s, $s \in \tilde{S}$

$H_s = \{b \in R^N : 0 \leq b_i \leq b_i^{max}, \forall i \in I_3^s; b_i = b_i^{max}, \forall i \in I_1^s; b_i = 0, \forall i \in I_2^s\}, s \in \tilde{S}$

**Proposition 1.** Consider arbitrary task $s, s \in \tilde{S}$, then necessary conditions for an extremum of the function $W_s$ in the point $b^s \in Int(F_s)$ contain following equations:

$$\forall i, j \in I_3^s : \frac{d\omega_i}{db_i}(b_i^s) = \frac{d\omega_j}{db_j}(b_j^s) = -\frac{V_s}{B_s}(b^s). \quad (12)$$

The proof of Proposition 1 is provided in Appendix B.

**Proposition 2.** Regard arbitrary task $s, s \in \tilde{S}$. Let $b^{s*} \in Int(F_s)$ and $b^{s*}$ is a stationary point of the function $W_s$. Then the point $b^{s*}$ represents the point of a local maximum of the function under consideration.

The proof of this result you can find in Appendix B.

**Remark 1.**

It is important to note that, in general, a function $W_s$ can be not concave on the set $F_s$. Consider this simple example, which elaborates describing situation:

$$\begin{cases} W_s = (b_1 + b_2) * [\left(1 - \frac{b_1^2}{4}\right) + (10 - 5 * b_2^2)] \\ F_s = \{(b_1, b_2) \in R^2 | 0 \leq b_1 \leq 2, 0 \leq b_2 \leq \sqrt{2}\} \end{cases}$$

It is easy to show, that this function is not concave on considering set (the figure below supports this fact graphically).

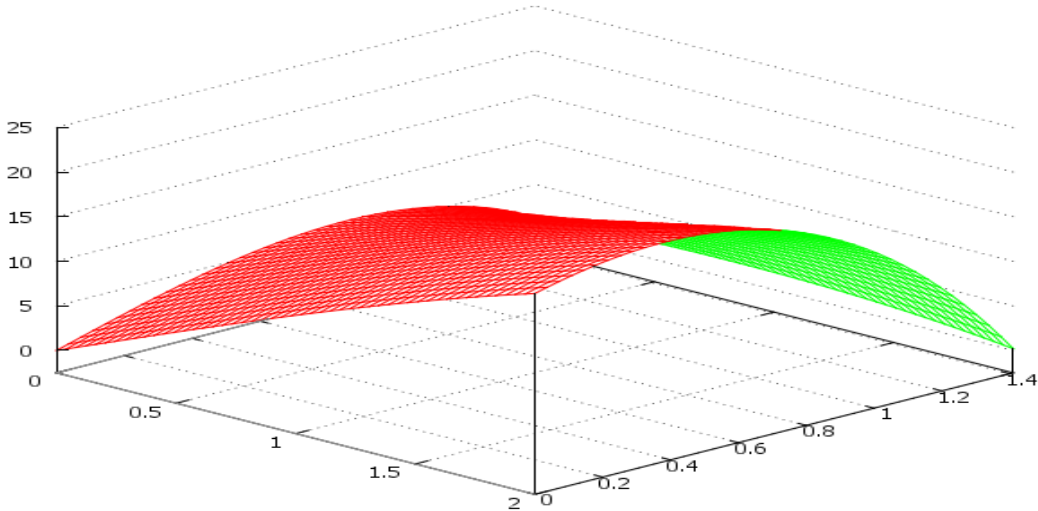

Now we are ready to formulate a statement, which describes main properties of the solution to the problem (8).

**Theorem 2.** Consider optimization problem (8), where all trade-off functions satisfy to the set of assumptions (9). For any task $s \in \tilde{S}$, denote:

$$X_s = \{b | b \in H_{\tilde{s}\tilde{s}}, b = (b_1, \ldots, b_N): b_i = b_i^{max}, \forall i \in I_1^s; b_i = 0, \forall i \in I_2^s; \forall i,j \in I_3^s: \frac{d\omega_i}{db_i}(b_i)$$

$$= \frac{d\omega}{db_j}(b_j) = -\frac{V_s(b)}{B_s(b)}\}.$$

$$Y = \{b \in R^N | b - \text{ the solution to the task (8)-(9)}\}.$$

We claim that following statement is true:

$$b^* \in Y \leftrightarrow b^* = \underset{b \in (\cup_{s \in \tilde{S}} X_s) \cup Vert(H_{\tilde{s}\tilde{s}})}{\operatorname{argmax}} W(b)$$

Proof of this theorem follows immediately from Proposition 1 and standard optimization techniques. Proposition 2 and Remark 1 announce that in the absence of additional information we cannot reduce the set of potential points of solution to the problem (8). So, in general, there are no points from the set $(\cup_{s \in \tilde{S}} X_s) \cup Vert(H_{\tilde{s}\tilde{s}})$ which can be precluded from our consideration as points where the solution to the problem (8) cannot be.

To conclude, we can make some inferences from this model:

1. In cases, when the differentiation of types of cells in a small – sized colony is essential, the colony tends to specialization. For example, if the diversity in cells is high, then conditions (12) may be not performed for each auxiliary task $s \in \tilde{S}, |I_3^s| \geq 2$, constructed for the chosen task (8)-(9), so the colony under consideration would exhibit complete specialization. However, note that high differentiation of cells types in small – sized colonies emerges rarely. It can be explained for some different reasons. Statistically, the possibility of appearance of essential gene mutation caused types differentiation in small – sized colony is lower than in large – sized colony.

2. In more general cases, when type's diversity in the small – sized colony is week, an unspecialized state may bring colony some benefits. However, these benefits may be only local and in optimum in the colony some cells would be specialized, others – unspecialized. The amount of specialized cells in small – sized colony depends on the level of type's diversity in this colony.

All in all, provided model predicts that colonies with low initial costs of reproduction would be organized as following: the majority of cells would be included in unspecialized layer, other cells may perform some functions: somatic or reproductive. These specialized cells emerge due to the type's diversity.

## 4. Full optimization problem with different types of cells and energy restriction: the case of convex (concave) trade-off functions

### 4.1. Formulation of the problem

According to [2], we can consider an energy restriction which has enough common form:

$$k_1 * B + k_2 * V \leq C, \text{ where } k_1 > 0, k_2 > 0, C > 0. \tag{11}$$

It shows that our colony is put into the environment with restricted amount of energy that should be divided between two basic components of fitness: viability and fecundity. This restriction reflects the fact that during the life history each individual existed in enough unfavorable habits. Moreover, we can suppose that these habits may significantly influence on the state of the colony under consideration. This is important for us, because determined restriction may change the decision of the colony about specialization in some cases.

So, we can represent the full optimization problem with different types of cells and energy restriction as follow:

$$\begin{cases} W(b) = \sum_{i=1}^{N} b_i * \sum_{i=1}^{N} \vartheta_i(b_i) \to max_b \\ k_1 * \sum_{i=1}^{N} b_i + k_2 * \sum_{i=1}^{N} \vartheta_i(b_i) \leq C, \\ \forall i = 1..N: 0 \leq b_i \leq b_i^{max}. \end{cases} \tag{12}$$

Also, we define the set of assumption which should be performed for convex (concave) trade-off functions:

1. for any $i = \overline{1,N}$ let $b_i^{max} \in R, 0 < b_i^{max} < \infty$;

2. $\forall i = \overline{1,N}: \vartheta_i: [0, b_i^{max}] \to R, \vartheta_i \in \mathbb{C}^2_{[0,b_i^{max}]}$;

3. $\forall i = \overline{1,N}: \vartheta_i(0) = v_i^{max}, 0 < v_i^{max} < \infty$;

4. $\forall i = \overline{1,N}: \vartheta_i(b_i^{max}) = 0$;

5. $\forall i = \overline{1,N}: \frac{d\vartheta_i(b_i)}{db_i} < 0, \forall b_i \in [0, b_i^{max}]$;

6. $\forall i = \overline{1,N}: \frac{d^2\vartheta_i(b_i)}{db_i^2} > 0, \forall b_i \in [0, b_i^{max}], \left(\forall i = \overline{1,N}: \frac{d^2\vartheta_i(b_i)}{db_i^2} < 0, \forall b_i \in [0, b_i^{max}]\right)$.

Properties (1)-(6) were introduced earlier for convex (concave) trade-off functions. Provided model does not prohibit cells to have different trade-off functions as well as in previous models.

Now, our aim is to find out the solution to the problem (12) and infer about the possibility of cell's specialization in this solution.

## 4.2. Analysis of the model

Now we provide common results which help us to understand more about the solution to the task (12). First of all, we investigate properties of the domain of the problem under consideration.

Denote:

$$A = \{b \in R^N, b = (b_1, \ldots, b_N) | \forall i = 1..N: 0 \leq b_i \leq b_i^{max}\}.$$

**Proposition 3.** Consider the domain D of the task (12), where all trade-off functions are convex (concave):

$$D = \{b \in R^N, b = (b_1, \ldots, b_N) | \forall i = 1..N: 0 \leq b_i \leq b_i^{max}; k_1 * \sum_{i=1}^{N} b_i + k_2 * \sum_{i=1}^{N} \vartheta_i(b_i) \leq C\};$$

Then, D (A\D) is a convex set.

The proof of this result you can find in Appendix C.

Further, suppose that all trade-off functions are convex (concave).

First of all, we should investigate the behavior of the function W in the energetic surface, so we should regard following task:

$$\begin{cases} W(b) = \sum_{i=1}^{N} b_i * \sum_{i=1}^{N} \vartheta_i(b_i) \to max_b \\ k_1 * \sum_{i=1}^{N} b_i + k_2 * \sum_{i=1}^{N} \vartheta_i(b_i) = C, \\ \forall i = 1..N: 0 \leq b_i \leq b_i^{max}. \end{cases} \quad (13)$$

Due to a comprehension about the solution to the problem (13), Theorem 1 and Theorem 2 we receive the solution to the task (12). That's why we begin with the task (13).

**Proposition 4.** Consider optimization problem (13). Consider following system of equations and inequities:

$$\begin{cases} \sum_{i=1}^{N} b_i = \frac{C}{2 * k_1}, \\ \sum_{i=1}^{N} \vartheta_i(b_i) = \frac{C}{2 * k_2}, \\ \forall i = 1..N: 0 \leq b_i \leq b_i^{max}. \end{cases} \quad (14)$$

Let a set of points which satisfy to (14) is not empty. Then these and only these points represent the solution to the problem (13).

The proof of this proposition is provided in Appendix C.

Further we investigate main properties of this solution and underlie cases when system from Proposition 4 represents not only the solution to the problem (13), but the solution to the problem (12) too. But now, in order to make our analysis of the problem (13) more precise, we provide a one more proposition.

Consider task (13). Choose arbitrary and fix parameters of this task according to assumptions (1)-(6). Further we would consider only this chosen problem as the problem (13). Because we chose this task arbitrary, all inferences about the solution to this problem would hold to each task (13) with any possible parameters satisfying to (1)-(6).

Next we would construct auxiliary optimization problems based on parameters of chosen task (13). These tasks help us to solve chosen optimization problem (13) and infer more about its solution.

**Definition 3.** Let $\Xi$ be the class of optimization problems. $\xi \in \Xi$ if and only if $\xi$ is an optimization problem which form is represented below:

$$\begin{cases} W_\xi = \left(\sum_{i \in I_1^\xi} b_i^{max} + \sum_{i \in I_3^\xi} b_i\right) * \left(\sum_{i \in I_2^\xi} v_i^{max} + \sum_{i \in I_3^\xi} \vartheta_i(b_i)\right) \to max_{b_i, i \in I_3^\xi} \\ k_1 * \left(\sum_{i \in I_1^\xi} b_i^{max} + \sum_{i \in I_3^\xi} b_i\right) + k_2 * \left(\sum_{i \in I_2^\xi} v_i^{max} + \sum_{i \in I_3^\xi} \vartheta_i(b_i)\right) = C. \end{cases} \quad (15)$$

$$\text{where} \begin{cases} I_1^\xi \subset \{1,..,N\}; I_2^\xi \subset \{1,..,N\}; I_3^\xi \subseteq \{1,..,N\}; \\ I_1^\xi \cup I_2^\xi \cup I_3^\xi = \{1,..,N\}; \\ I_1^\xi \cap I_2^\xi = I_1^\xi \cap I_3^\xi = I_2^\xi \cap I_3^\xi; \\ \left|I_3^\xi\right| \geq 1; \end{cases} \quad (16)$$

So, any task $\xi$ from the class $\Xi$ has a form (15) and is determined exactly of a triple of sets $\left(I_1^\xi, I_2^\xi, I_3^\xi\right)$ satisfying to conditions (16).

**Proposition 5.** Consider optimization task (13). Suppose that the solution to system (14) is empty. Let $b^*$ represents the solution to the problem (13). Then following statement about $b^*$ is true:

$$b^* \in \left\{(b_1^*, ..., b_N^*) \in A \cap M_1 \;\middle|\; \begin{array}{l} \exists \xi \in \Xi : [\forall i \in I_1^\xi : b_i^* = b_i^{max}, \forall i \in I_2^\xi : b_i^* = 0, \\ \forall i, j \in I_3^\xi : \frac{d\vartheta_i}{db_i}(b_i^*) = \frac{d\vartheta_j}{db_j}(b_j^*)]. \end{array} \right\} \quad (17)$$

The proof of this proposition is provided in Appendix C.

In general, we describe the behavior of fitness function on the energetic surface. Now, due to Theorem 1, Theorem 2, Proposition 4 and Proposition 5, we can infer more about the solution to the problem (12).

Furthermore, we provide following result about the solution to the problem (12):

**Proposition 6.** Consider optimization problem (12). Let the system (14) is not empty. Regard all points that satisfy to the system (14). We claim that these and only these points represent the solution to the problem (12).

The proof of this proposition you can find in Appendix C.

We have provided a set of proposition, which can help us to understand more about the solution to the problem under consideration. Now due to these propositions we can make some inferences:

- Proposition 3 allows us to conclude that points, described in Proposition 6, create a finite number of connected components. This fact means that suggested model is partially robust and can be completely robust.

- Due to Proposition 6, in optimum, colony with different types of cells and energy restriction may be indifferent between some optimal patterns of states. Arbitrary chosen cell may be soma or germ in some states or may be unspecialized in other.

- Consider each optimal pattern: a small change in unobserved parameters of the model, which does not lead to leaving this pattern, tend to small and continuous change in optimal state. It means that the model is robust within each pattern.

- In each optimal pattern levels of fecundity and viability of each cell lies in limited ranges, which are individual for each cell. This result reflects the fact that some cell in the colony may lose the potential ability to achieve, for example, high level of fecundity, but does not lose the possibility to perform a reproductive function at all. Describing situation represents the intermediate between unspecialized colonies and full-specialized multicellular organisms.

- We provide some intuition to solutions revealed in Proposition 6. Consider optimal pattern. Suppose that initially there is an external irritant impacted on some cells from the colony, so the solution to the problem under consideration belongs to this pattern. Further, suppose that this irritant change its location a bit, so the impact of this irritant on noted cells reduced a bit, but impact on other cells increased a bit (all in all, this displacement of the irritant does not change the total amount of energy from restriction). Solution revealed in Proposition 6 allows colony to reallocate fecundity and viability between cells in response to changing in, for instance, location of irritant, such that the levels of fecundity and viability of a colony remain constant.

- Finally, we discuss Proposition 5, which provides the form of necessary conditions of the extremum of the task (13). Moreover points revealed in Proposition 5 can be points of solution to problem (12). Generally speaking, this result means that in some cases the solution to problem (12) represents only one point, not a set of patterns or one pattern. Thus, in some cases, in optimum each cell of the colony should perform strictly determined function and small deviation in unobservable parameters does not change this solution. Furthermore, the level of specialization in the colony in the optimal point is determined by the ratio included first derivatives of trade-off functions. And this one is influenced by the level of type's differentiation. For example, if the diversity in types in the colony is high (large – sized colonies), this colony is more predisposed to specialization than another one with small level of type's diversity (small – sized colonies).

## 5. Conclusion.

We developed models provided in [1], [2], and attempt to get robust models investigated how differentiation of types and energy constraints influence on the optimal behavior of colonies with different size (so, different initial costs of reproduction).

Constructed model show that each large sized colony with high initial costs of reproduction tends to full specialization, no matter are all cells in this colony identical or are there cells with different types in this colony. Moreover, in this model not exactly the half of cells from the colony should specialize in, for example, soma. The level of type's diversity determines the number of cells specialized in soma.

In small – sized colonies with low initial costs of reproduction, when type's diversity is week, an unspecialized state may bring colony some benefits. However, these benefits may be only local and in optimum in the colony some cells would be specialized, others – unspecialized. The amount of specialized cells in small – sized colony depends on the level of type's diversity in this colony.

Adding energy constraint, we may receive robust models even in convex case. Generally speaking, these models would be partially robust. In optimum, the colony with different types of cells and energy restriction may be indifferent between some optimal patterns of states. Arbitrary chosen cell may be soma or germ in some states or may be unspecialized in other. These patterns embody the idea of robustness.

All in all, in this work we elaborate models, provided in [2] and regard cases with more general suggestions about curvatures of intrinsic trade-off functions of all cells in the colony under consideration.

## Appendix A.

**Proof of Theorem 1.** First of all, consider a lemma.

**Lemma1.** For each $p \in \tilde{\mathcal{P}}, |I_3^p| > 1$: let $b_p^* \in IntF_p$, then $b_p^*$ cannot be a solution to the optimization problem p.

**Proof.** Choose any $p \in \tilde{\mathcal{P}}, |I_3^p| > 1$. Let $b_p^* \in IntF_p$. Suppose, $b_p^*$ - is a solution to the optimization problem p.

1). $b_p^* \in IntF_p \rightarrow \exists U_\varepsilon(b_p^*) \subset F_p$;

2). Consider a hyperplane $A = \{b \in Aff(F_p): \sum_{i \in I_3^p} b_i = \sum_{i \in I_3^p} b_{pi}^*\}$ and choose $x, y \in A \cap U_\varepsilon(b_p^*)$ such that $b_p^* = \mu * x + (1 - \mu) * y, \mu \in (0, 1)$;

3). According to the fact that $b_p^*$ represents the solution to the task p, $W(b_p^*) \geq \max\{W(x); W(y)\}$;

4). $W(b_p^*) - \mu * W(x) - (1-\mu) * W(y) = W(\mu * x + (1-\mu) * y) - \mu * W(x) - (1-\mu) * W(y) = B * \left(\sum_{i \in I_2^p} v_i^{max} + \sum_{i \in I_3^p} \varphi_i(\mu * x_i + (1-\mu) * y_i)\right) - \mu * W(x) - (1-\mu) * W(y) < B * \left(\sum_{i \in I_2^p} v_i^{max} + \mu * \sum_{i \in I_3^p} \varphi_i(x_i) + (1-\mu) * \sum_{i \in I_3^p} \varphi_i(y_i)\right) - \mu * B * \left(\sum_{i \in I_2^p} v_i^{max} + \sum_{i \in I_3^p} \varphi_i(x_i)\right) - (1-\mu) * B * \left(\sum_{i \in I_2^p} v_i^{max} + \sum_{i \in I_3^p} \varphi_i(y_i)\right)$

Denote $B = \sum_{i \in I_1^p} b_i^{max} + b$; $b = \sum_{i \in I_3^p} x_i = \sum_{i \in I_3^p} y_i$; $e = \sum_{i \in I_3^p} \varphi_i(x_i)$; $f = \sum_{i \in I_3^p} \varphi_i(y_i)$;

Thus: $W(b_p^*) - \mu * W(x) - (1-\mu) * W(y) < B * \left(\sum_{i \in I_2^p} v_i^{max} + \mu * e + (1-\mu) * f\right) - \mu * B * \left(\sum_{i \in I_2^p} v_i^{max} + e\right) - (1-\mu) * B * \left(\sum_{i \in I_2^p} v_i^{max} + f\right) = 0 \to W(b_p^*) < \mu * W(x) + (1-\mu) * W(y)$

5). $\text{Max}\{W(x); W(y)\} \le W(b_p^*) < \mu * W(x) + (1-\mu) * W(y) \le \max\{W(x); W(y)\} \to$

$$\to \max\{W(x); W(y)\} < \max\{W(x); W(y)\};$$

So, we have a contradiction. It means that the lemma is proved.

Now we start to prove Theorem 1. View optimization problem (4).

1. If N=1, then, obviously, the statement (7) is true.

2. Suppose, N>1. Then, optimization problem (4) belongs to the class $\widetilde{\mathcal{P}}$. Consider point $b \in Int(H_{\tilde{p}\tilde{p}})$, (note, that $H_{\tilde{p}\tilde{p}} = F_{\tilde{p}\tilde{p}}$). According to the Lemma 1, b cannot be a solution to the optimization problem (4), so the solution to (4) belongs to $\partial H_{\tilde{p}\tilde{p}}$.

Consider the behavior of function W in $\partial H_{\tilde{p}\tilde{p}}$. Choose one variable and charge it the value 0 or $b_i^{max}$. Thus we describe a set $\partial H_{\tilde{p}\tilde{p}} = \bigcup_{p \in \tilde{\mathcal{P}}: |I_3^p|=N-1} H_{\tilde{\mathcal{P}}}$. So, we get tasks, each of them would be equivalent to some $p \in \widetilde{\mathcal{P}}: |I_3^p| = N - 1$. According to Lemma 1, for all $b \in ReInt(H_{\tilde{p}})$, it follows, that b is not a solution to a problem (4). So, we should find the solution to (4) in a set $\bigcup_{p \in \tilde{\mathcal{P}}: |I_3^p|=N-2} H_{\tilde{p}}$, where our conclusions are the similar. So, we should repeat our procedure until we get tasks in edges and vertexes of $H_{\tilde{p}\tilde{p}}$. The solution to (4) belongs to vertexes or edges of hyper parallelepiped $H_{\tilde{p}\tilde{p}}$, so:

$$(b^* \in Vert(H_{\tilde{p}\tilde{p}})) \vee (b^* \in F), \text{Q. E. D.}$$

## Appendix B

**Proof of Proposition 1.** This proposition is obvious and follows immediately from usual optimization techniques. Consider arbitrary task $s, s \in \tilde{S}$. Let $b^s \in Int(F_s)$ represents the point of extremum of the function $W_s$. Then following equation should be satisfied in the point $b^s$:

$$\nabla W_s(b^s) = 0 \leftrightarrow \left(\sum_{i \in I_2^s} v_i^{max} + \sum_{i \in I_3^s} \omega_i(b_i^s)\right) + \left(\sum_{i \in I_1^s} b_i^{max} + \sum_{i \in I_3^s} b_i^s\right) * \frac{d\omega_j}{db_j}(b_j^s) = 0, \forall j \in I_3^s$$

$$V_s(b^s) + B_s(b^s) * \frac{d\omega_j}{db_j}(b_j^s) = 0, \forall j \in I_3^s$$

$$\forall i,j \in I_3^s: \frac{d\omega_i}{db_i}(b_i^s) = \frac{d\omega_j}{db_j}(b_j^s) = -\frac{V_s}{B_s}(b^s);$$

So, the required result is proved.

**Proof of Proposition 2.** Choose any $s \in \tilde{S}$. Consider any $b^{s*} \in Int(F_s)$, representing a stationary point of the function $W_s$. According to previous proposition, the following statement is true:

$$\forall i,j \in I_3^s: \frac{d\omega_i}{db_i}(b_i^{s*}) = \frac{d\omega_j}{db_j}(b_j^{s*}) = -\frac{V_s}{B_s}(b^{s*});$$

In order to research the behavior of regarding function in this point, it is necessary to use second order conditions.

$$\frac{\partial^2 W_s}{\partial b_i^2}(b^{s*}) = 2 * \frac{d\omega_i}{db_i}(b_i^{s*}) + B_s(b^{s*}) * \frac{d^2\omega_i}{db_i^2}(b_i^{s*}), \forall i \in I_3^s;$$

Let $g \in I_3^s$, then:

$$\frac{\partial^2 W_s}{\partial b_i^2}(b^{s*}) = 2 * \frac{d\omega_g}{db_g}(b_g^{s*}) + B_s(b^{s*}) * \frac{d^2\omega_i}{db_i^2}(b_i^{s*}), \forall i \in I_3^s;$$

$$\frac{\partial^2 W_s}{\partial b_i b_j}(b^{s*}) = \frac{d\omega_j}{db_j}(b_j^{s*}) + \frac{d\omega_i}{db_i}(b_i^{s*}), \forall i \neq j \in I_3^s;$$

$$\frac{\partial^2 W_s}{\partial b_i b_j}(b^{s*}) = 2 * \frac{d\omega_g}{db_g}(b_g^{s*}), \forall i \neq j \in I_3^s;$$

Introduce some useful redesignation:

$$\psi_i = \frac{d^2\omega_i}{db_i^2}(b_i^{s*}), \forall i \in I_3^s;$$

$$\theta = 2 * \frac{d\omega_g}{db_g}(b_g^{s*});$$

$$B_s(b^{s*}) = Z;$$

$$t_1^s = \min\{i | i \in I_3^s\}; \ t_d^s = \min\{i | i \in I_3^s \setminus \bigcup_{f=1}^{d-1}\{t_f^s\}\}, \forall d \in \{2,..,|I_3^s|\};$$

Then, we can represent the Hesse matrix, calculating in the point $b^{s*}$ as following:

$$He(s, b^{s*}) = \begin{pmatrix} Z * \psi_{t_1^s} + \theta & \theta & \cdots & \theta \\ \theta & Z * \psi_{t_2^s} + \theta & \cdots & \theta \\ \vdots & \vdots & \ddots & \vdots \\ \theta & \theta & \cdots & Z * \psi_{t_{|I_3^s|}^s} + \theta \end{pmatrix}$$

**Lemma 2.** Consider following determinant.

$$I(n) = \begin{vmatrix} \tilde{Z}*\tilde{\psi}_1 + \tilde{\theta} & \tilde{\theta} & \cdots & \tilde{\theta} \\ \tilde{\theta} & \tilde{Z}*\tilde{\psi}_2 + \tilde{\theta} & \cdots & \tilde{\theta} \\ \vdots & \vdots & \ddots & \vdots \\ \tilde{\theta} & \tilde{\theta} & \cdots & \tilde{Z}*\tilde{\psi}_n + \tilde{\theta} \end{vmatrix}.$$

Where $\tilde{\theta}, \tilde{Z}, \tilde{\psi}_1, \ldots, \tilde{\psi}_n \in R, n \geq 1$. Then:

$$I(n) = \tilde{Z}^{n-1} * \left[\tilde{Z} * \prod_{i=1}^{n} \tilde{\psi}_i + \tilde{\theta} * \sum_{k \in K} \prod_{i \in k} \tilde{\psi}_i \right];$$

Where $K = \{k \in 2^{\{1,\ldots,n\}} | |k| = n-1\}$.

**Proof.** Denote:

$$a_i = (0 \quad \ldots \quad \tilde{Z}*\tilde{\psi}_i \quad \ldots \quad 0)^T, \forall i = 1,\ldots,n; \; \Theta = (\tilde{\theta}, \quad \ldots \quad \tilde{\theta}, \quad \ldots \quad \tilde{\theta},)^T;$$

This denotation simplifies the representation of viewed determinant and allows us to calculate the determinant easily.

$$I(n) = |a_1 + \Theta; a_2 + \Theta; \ldots; a_n + \Theta| = |a_1; a_2 + \Theta; \ldots; a_n + \Theta| + |\Theta; a_2 + \Theta; \ldots; a_n + \Theta|;$$

$$I(n) = |a_1; a_2; \ldots; a_n + \Theta| + |a_1; \Theta; \ldots; a_n| + |\Theta; a_2; \ldots; a_n|;$$

$$I(n) = |a_1; a_2; \ldots; a_n| + |\Theta; a_2; \ldots; a_n| + |a_1; \Theta; \ldots; a_n| + \cdots + |a_1; a_2; \ldots; \Theta|;$$

$$|a_1; a_2; \ldots; a_n| = \tilde{Z}^n * \prod_{i=1}^{n} \tilde{\psi}_i;$$

$$|a_1; \ldots; \Theta; \ldots; a_n| = \begin{vmatrix} \tilde{Z}*\tilde{\psi}_1 & \ldots & \tilde{\theta} \ldots & 0 \\ 0 & \ldots & \tilde{\theta} \ldots & 0 \\ \vdots & \vdots & \vdots \ldots & \vdots \\ 0 & \ldots & \tilde{\theta} \ldots & \tilde{Z}*\tilde{\psi}_n \end{vmatrix} = \tilde{\theta}*(-1)^{2*i}*\tilde{Z}^{n-1}*\prod_{j=1,j\neq i}^{n} \tilde{\psi}_i$$

$$= \tilde{\theta} * \tilde{Z}^{n-1} * \prod_{j=1,j\neq i}^{n} \tilde{\psi}_i;$$

Finally, we conclude that:

$$I(n) = \tilde{Z}^{n-1} * \left[\tilde{Z} * \prod_{i=1}^{n} \tilde{\psi}_i + \tilde{\theta} * \sum_{k \in K} \prod_{i \in k} \tilde{\psi}_i \right].$$

Where $K = \{k \in 2^{\{1,\ldots,n\}} | |k| = n-1\}$. So, Lemma 2 is proved.

It is obvious that all corner minors of matrix $He(s, b^{s*})$ have the form of $I(1), \ldots, I(|I_3^s|)$ accordingly where holds following equations:

$$\begin{cases} \tilde{\psi}_i = \psi_{t_i^s}, \forall i = 1,\ldots,|I_3^s|; \\ \tilde{Z} = Z; \\ \tilde{\theta} = \theta. \end{cases}$$

Now we can apply a Sylvester criterion in order to determine the type of a stationary point. Note that:

$$\begin{cases} \psi_{t_i^s} < 0, \forall i = 1,\ldots,|I_3^s|; \\ \theta < 0; \\ Z > 0. \end{cases}$$

We can see that all even corner minors of the Hesse matrix are positive and all odd corner minors are negative. Thus we can conclude that viewed Hesse matrix has negative definiteness.

That's why, $b^{s*}$ is the point of a local maximum of the function $W_s$.

Proposition 2 is proved.

**Appendix C.**

**Proof of Proposition 3.** Suppose, for instance, that all trade-off functions are convex. Introduce sets A and M:

$$A = \{b \in R^N, b = (b_1, \ldots, b_N) | \forall i = 1..N: 0 \leq b_i \leq b_i^{max}\};$$

$$M = \{b \in R^N, b = (b_1, \ldots, b_N) | k_1 * \sum_{i=1}^N b_i + k_2 * \sum_{i=1}^N \vartheta_i(b_i) \leq C\}.$$

$$\text{Let } x, y \in M \rightarrow \begin{cases} k_1 * \sum_{i=1}^N x_i + k_2 * \sum_{i=1}^N \vartheta_i(x_i) \leq C \\ k_1 * \sum_{i=1}^N y_i + k_2 * \sum_{i=1}^N \vartheta_i(y_i) \leq C \end{cases};$$

$$\text{Consider } z = \lambda * x + (1 - \lambda) * y; k_1 * \sum_{i=1}^N z_i + k_2 * \sum_{i=1}^N \vartheta_i(z_i)$$

$$< \lambda * \left[ k_1 * \sum_{i=1}^N x_i + k_2 * \sum_{i=1}^N \vartheta_i(x_i) \right] + (1 - \lambda) * \left[ k_1 * \sum_{i=1}^N x_i + k_2 * \sum_{i=1}^N \vartheta_i(x_i) \right]$$

$$= C \rightarrow k_1 * \sum_{i=1}^N z_i + k_2 * \sum_{i=1}^N \vartheta_i(z_i) < C \rightarrow z \in M \rightarrow M = Conv(M).$$

$$M = Conv(M); A = Conv(A); D = A \cap M \rightarrow D = Conv(D).$$

So, the proposition is proved.

**Proof of Proposition 4.** Introduce following denotation:

$$E = \{b \in R^N, b = (b_1, \ldots, b_N) | \sum_{i=1}^{N} b_i = \frac{C}{2 * k_1}\};$$

$$M_1 = \{b \in R^N, b = (b_1, \ldots, b_N) | k_1 * \sum_{i=1}^{N} b_i + k_2 * \sum_{i=1}^{N} \vartheta_i(b_i) = C\};$$

$$Z: R^N \to R: Z(b) = \frac{1}{k_2} * B(b) * (C - k_1 * B(b)).$$

We can infer following results:

$$\forall b \in M_1: W(b) = Z(b) \to \begin{cases} W(b) \to max_b \\ b \in M_1 \end{cases} \leftrightarrow \begin{cases} Z(b) \to max_b \\ b \in M_1 \end{cases}.$$

$$(b^* \in argmax_{b \in R^N} Z(b)) \leftrightarrow (b^* \in E) - \text{due to the concavity of function } Z(b).$$

Let $b^* \in E$. If $b^* \in M_1$, then $b^* \in argmax_{b \in M_1} W(b)$.

Let $b^* \in E \cap M_1 \cap A$ ($b^*$ satisfies (14)), then $b^* \in argmax_{b \in M_1 \cap A} W(b)$.

Let $b^{**} \in M_1 \cap A$ (the domain to the problem (13)), but $b^{**}$ does not satisfy to (14).

Then $b^{**} \notin E$. So, $Z(b^{**}) < Z(b^*) \to W(b^{**}) < W(b^*)$.

Last statement means that the point $b^{**}$ can not represent the solution to the problem (13).

So, the proposition is proved.

**Proof of Proposition 5.** Regard task (13). It requires maximizing fitness function of the colony on the segment of energy surface, which is enclosed in N-dimensional hyper parallelepiped. According to the standard techniques, we should to regard separate tasks: the first task - find points of conditional extremum of function under consideration on energetic restriction belonging to the interior of the hyper parallelepiped; second part of tasks - find points of conditional extremum of function under consideration on energetic restriction belonging to relative interiors of each plane and edges of hyper parallelepiped under consideration. Note, that these reduced tasks are equivalent (bijectively) to corresponding tasks from the set $\Xi$. All in all, in order to figure out points of solution to the problem (13) we should examine all stationary points of each task $\xi \in \Xi$, belonging to corresponding plane of the hyper parallelepiped (more precisely, to the image of corresponding plane, received due to bijection).

Chose and fix arbitrary task $\xi \in \Xi$ and investigate all points from corresponding plane of hyper parallelepiped in which conditional extremum of this task is possible. According to Lagrange method we should construct additional function $L_\xi$.

$$L_\xi = B_\xi * V_\xi - \lambda * [k_1 * B_\xi + k_2 * V_\xi - C].$$

First of all, highlight an extraordinary case:

$$k_1 + k_2 * \frac{d\vartheta_i}{db_i}(b_i) = 0, \forall i \in I_3{}^\xi \to \frac{d\vartheta_i}{db_i}(b_i) = \frac{d\vartheta_j}{db_j}(b_j) = -\frac{k_1}{k_2}, \forall i, j \in I_3{}^\xi.$$

Then we provide ordinary case and necessary conditions for conditional extremum of function $W_\xi$ on corresponding energetic restriction $k_1 * B_\xi + k_2 * V_\xi = C$:

$$\frac{\partial L_\xi}{\partial b_i} = V_\xi + B_\xi * \frac{d\vartheta_i}{db_i}(b_i) - \lambda * \left[k_1 + k_2 * \frac{d\vartheta_i}{db_i}(b_i)\right] = 0, \forall i \in I_3{}^\xi;$$

Note that due to the fact that the solution to system (14) is empty, we can infer following results:

$$\begin{cases} \lambda * k_1 - V_\xi \neq 0 \\ B_\xi - \lambda * k_2 \neq 0 \end{cases}$$

So, we can write:

$$\frac{d\vartheta_i}{db_i}(b_i) = \frac{\lambda * k_1 - V_\xi}{B_\xi - \lambda * k_2}, \forall i \in I_3{}^\xi;$$

$$\frac{d\vartheta_i}{db_i}(b_i) = \frac{d\vartheta_j}{db_j}(b_j) = \frac{\lambda * k_1 - V_\xi}{B_\xi - \lambda * k_2}, \forall i, j \in I_3{}^\xi.$$

Applying the bijection, we can find that each point in which solution to the problem (13) can be (if the solution to the problem (14) is empty) should belong to following set:

$$\left\{(b_1^*, \ldots, b_N^*) \in A \cap M_1 \middle| \begin{array}{l} \exists \xi \in \Xi : [\forall i \in I_1{}^\xi : b_i^* = b_i^{max}, \forall i \in I_2{}^\xi : b_i^* = 0, \\ \forall i, j \in I_3{}^\xi : \frac{d\vartheta_i}{db_i}(b_i^*) = \frac{d\vartheta_j}{db_j}(b_j^*)]. \end{array}\right\}$$

Proposition 5 is proved.

**Proof of Proposition 6.** Consider following set:

$$\varkappa = \{b \in R^N | \ b \text{ satisfies to the set (14)}\}.$$

1. Under hypothesis of proposition under consideration, this set is not empty.

2. Note that values of fitness function in each point from the set $\varkappa$ are equal to $\frac{C^2}{4k_1k_2}$.

3. Chose any two points $b^*$ and $b^{**}$ such that $b^* \in \varkappa, b^{**} \in D \backslash \varkappa$. We should show that $W(b^*) > W(b^{**})$.

4. Consider two possible cases:

- $b^{**} \in (A \cap M_1) \backslash \varkappa \to W(b^*) > W(b^{**})$ (Due to the Proposition 4).

- $b^{**} \in D \backslash (A \cap M_1)$. Then: $k_1 * \sum_{i=1}^{N} b_i{}^{**} + k_2 * \sum_{i=1}^{N} \vartheta_i(b_i{}^{**}) = \sigma < C$. Consider following set and a function:

$$M_2 = \{b \in R^N, b = (b_1, \ldots, b_N) | k_1 * \sum_{i=1}^{N} b_i + k_2 * \sum_{i=1}^{N} \vartheta_i(b_i) = \sigma\};$$

$$Z_1: R^N \to R: Z_1(b) = \frac{1}{k_2} * B(b) * (\sigma - k_1 * B(b)).$$

We can note that following inequity is true:

$$\forall b \in R^N: Z_1(b) \leq \frac{\sigma^2}{4k_1 k_2}.$$

Moreover, we figure out:

$$\forall b \in M_2: Z_1(b) = W(b) \to W(b^{**}) = Z_1(b^{**}) \leq \frac{\sigma^2}{4k_1 k_2} < \frac{C^2}{4k_1 k_2} = W(b^*)$$

So, $W(b^*) > W(b^{**})$ in this case too.

The proposition is proved.

## Acknowledgments

The author thank DeCAn laboratory of NRU HSE for partial financial support. The author is grateful to Professor F. Aleskerov for plenty enlightening discussions and help.

## References

1. Richard E. Michod, Yannick Viossat, Cristian A. Solari, Mathilde Hurand, Aurora M. Nedelcu. Life-history evolution and the origin of multicellularity, Journal of Theoretical Biology, Volume 239, Issue 2, 21 March 2006, P. 257–272.

2. F. Aleskerov, D. Tverskoy. Life history evolution and the origin of multicellularity: the case of different types of cells, Working Paper WP7/2014/05, Series WP7, Mathematical methods for decision making in economics, business and politics, NRU HSE.

3. Cristian A. Solari, John O. Kessler, Raymond E. Goldstein. Motility, mixing, and multicellularity, Genetic Programming and Evolvable Machines, Volume 8, Issue 2, June 2007, P. 115–129.

4. Claus Rueffler, Joachim Hermisson, and Gunter P. Wagner. Evolution of functional specialization and division of labor, PNAS, Volume 109, Issue 6, 7 February 2012, P. E326–E335.

5. Walter Bossert, Chloe X. Qi, John A. Weymark. Extensive social choice and the measurement of group fitness in biological hierarchies, Biology and Philosophy, 28, 2013, P. 75–98.

6. Cristian A. Solari, Aurora M. Nedelcu, and Richard E. Michod. Fitness and Complexity in Volvocalean Green Algae, AAAI Technical Report SS-03-02, 2003.